\documentclass[10pt,journal]{IEEEtran}
\usepackage{amsmath,amsfonts}
\usepackage{algorithmic}
\usepackage{algorithm}
\usepackage{array}
\usepackage{textcomp}
\usepackage{stfloats}
\usepackage{url}
\usepackage{verbatim}
\usepackage{graphicx}
\usepackage{multicol}
\usepackage{subfig}
\usepackage{cite}
\usepackage{comment}
\usepackage{multirow}
\usepackage{makecell} 
\usepackage{tabularx} 
\newcommand{\RNum}[1]{\uppercase\expandafter{\romannumeral #1\relax}}

\usepackage[dvipsnames]{xcolor}

\usepackage[colorlinks,allcolors=blue,bookmarks=false,hypertexnames=true]{hyperref} 

\begin{document}

\title{Redefining Radar Segmentation: Simultaneous Static-Moving Segmentation and Ego-Motion Estimation using Radar Point Clouds}

\author{\IEEEauthorblockN{Simin Zhu, Satish Ravindran, Alexander Yarovoy, \IEEEmembership{Fellow, IEEE}}, Francesco Fioranelli, \IEEEmembership{Senior Member, IEEE}

\thanks{Simin Zhu, Francesco Fioranelli, Alexander Yarovoy are with the Microwave Sensing, Signals and Systems (MS3) group, Delft University of Technology, 2628 CD, Delft, The Netherlands (e-mail: s.zhu-2@tudelft.nl; f.fioranelli@tudelft.nl; a.yarovoy@tudelft.nl)\par
Satish Ravindran is with NXP Semiconductors, San Jose, CA, USA (e-mail:satish.ravindran@nxp.com)}}

\markboth{Journal of \LaTeX\ Class Files,~Vol.~14, No.~8, December~2022}%
{Shell \MakeLowercase{\textit{et al.}}: A Sample Article Using IEEEtran.cls for IEEE Journals}


\maketitle

\begin{abstract}
Conventional radar segmentation research has typically focused on learning category labels for different moving objects. Although fundamental differences between radar and optical sensors lead to differences in the reliability of predicting accurate and consistent category labels, a review of common radar perception tasks in automotive reveals that determining whether an object is moving or static is a prerequisite for most tasks. To fill this gap, this study proposes a neural network-based solution that can simultaneously segment static and moving objects from radar point clouds. Furthermore, since the measured radial velocity of static objects is correlated with the motion of the radar, this approach can also estimate the instantaneous 2D velocity of the moving platform/vehicle (ego-motion). However, despite performing dual tasks, the proposed method employs very simple yet effective building blocks for feature extraction: multi-layer perceptrons (MLPs) and recurrent neural networks (RNNs). In addition to being the first of its kind in the literature, the proposed method also demonstrates the feasibility of extracting the information required for the dual task directly from unprocessed point clouds, without the need for cloud aggregation, Doppler compensation, motion compensation, or any other intermediate signal processing steps. To measure its performance, this study introduces a set of novel evaluation metrics and tests the proposed method using a challenging real-world radar dataset, RadarScenes. The results show that the proposed method not only performs well on the dual tasks, but also has broad application potential in other radar perception tasks. More qualitative results can be viewed here: \textcolor{blue}{https://youtu.be/3ejS1chSvQ8?si=uGRugVA63BCyvNBV}.
\end{abstract}


\section{Introduction}\label{introduction}
\IEEEPARstart{O}{ver} the past decade, the automotive industry has made tremendous progress in autonomous driving technology, revolutionizing today’s smart vehicles and transportation. As the goal shifts from testing to real-world driving environments, developing reliable sensor perception systems to ensure the safety of autonomous driving systems has become imperative. Common sensors used in perception systems include cameras, automotive radar, lidar, and sonar \cite{yang2025review}. Among these sensor options, autonomous radars play a vital role in providing robust perception information and demonstrate unparalleled advantages in the following aspects: firstly, the performance of radar perception is very robust to low light and adverse weather conditions such as rain, snow, and fog \cite{hong2022radarslam}; secondly, automotive radar can detect objects that are partially or completely obscured or even out of the line-of-sight of the radar \cite{he2024see}; thirdly, automotive radar can measure the radial velocity of detected objects, which can be used directly to estimate the speed of the ego-vehicle \cite{kellner2013instantaneous} and other moving objects \cite{kellner2013lateral}. 

\begin{figure}[!t]
\centering
\includegraphics[width=0.4\textwidth]{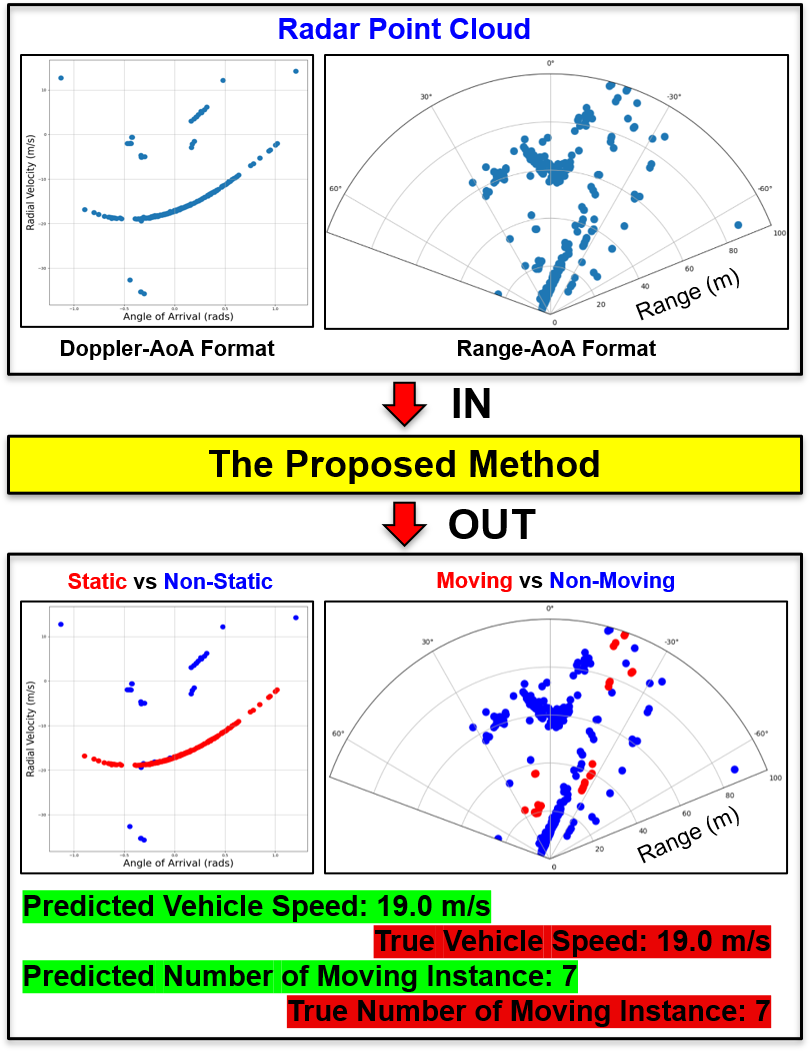}
\caption{The proposed method takes multidimensional radar point clouds as input, uses neural networks (NNs) for automatic feature extraction, and then segments static and moving objects. Based on the measured radial velocity of static objects, the method can estimate the ego-motion of the moving vehicle. The distinct moving instances can also be generated after applying a clustering algorithm to the predicted moving objects. In this example, the RadarScenes \cite{radar_scenes_dataset} dataset is used for testing.}
\label{fig:model_overview}
\end{figure}

The above advantages make automotive radar a powerful perception sensor in the automotive industry. Among many perception tasks, radar-based segmentation has gained significant attention in the past few years. The main objective in radar segmentation is to assign a class label to each point in the radar point cloud \cite{schumann2018semantic} or each cell in the radar data cube \cite{ouaknine2021multi}. The radar point cloud is generated after applying a detector such as one of the Constant False Alarm Rate (CFAR) algorithms \cite{rohling2007radar} on the radar data cube. Both data formats capture information such as range, radial velocity, and angle of arrival (AoA) of objects in the scanned environment. Therefore, performing segmentation on these data is crucial for scene understanding and driving safety. In the related radar literature, three categories of segmentation tasks have been explored, namely, semantic segmentation \cite{schumann2018semantic,schumann2019scene,ouaknine2021multi,zeller2022gaussian,fent2023radargnn,zhang2023spatial,wu2024mask,zhang2024tarss}, instance segmentation \cite{liu2022deep,xiong2022contrastive,zeller2023radar}, and panoptic segmentation \cite{zeller2024semrafiner}. Undoubtedly, these studies have demonstrated the great potential of radar sensors and laid a solid foundation for future perception systems.\par

However, most segmentation works have focused only on moving objects, while radar point clouds typically contain detections from moving objects (e.g., cars), static objects (e.g., buildings), and false positives (e.g., unidentified objects and multipath reflections). Further literature review shows that in other radar-based perception tasks, identifying which objects are static is also very important \cite{kellner2014instantaneous,grebner2023self,li2018high,holder2019real,giese2017road}. To locate static detections, some studies have to assume a static environment \cite{izquierdo2018multi,checchin2010radar}, while others usually rely on knowing the vehicle’s ego-motion \cite{grebner2023self,giese2017road} or use random sampling techniques \cite{kellner2014instantaneous,holder2019real} such as the Random Sample Consensus (RANSAC) algorithm \cite{fischler1981random}. While these remedies can help identify static objects, they either require external odometry sensors or assume that most objects are static, and often leave moving objects mixed with false positives, requiring further separation. \par

Therefore, to bridge the gap between the unprocessed radar point cloud and various perception applications that either require knowing the positions of static or moving objects, or the vehicle's speed, this study redefines the objective in conventional radar segmentation tasks. For this, it proposes a unified solution that can simultaneously perform the dual task of static and moving object segmentation and vehicle ego-motion estimation, as illustrated in Figure \ref{fig:model_overview}. To the best of our knowledge, this method is the first attempt to enable this dual task, and the results demonstrate that raw radar point clouds contain sufficient information to achieve both. In addition to this primary contribution, the proposed method offers the following advancements: \par

\begin{enumerate}

\item \textbf{Radar-only:} Unlike many other studies, the proposed method performs the dual tasks using only radar data. For example, it eliminates the need for odometry sensors to measure vehicle ego-motion to assist with radial velocity compensation or motion compensation. This preserves sensor independence and removes concerns about errors introduced by sensor synchronization or output glitches.

\item \textbf{No Aggregation:} The proposed method can handle sparse radar point clouds and does not require cloud aggregation from multiple radars or radar frames. Instead, to extract temporal features, the proposed method uses a moving window and takes multiple radar point clouds as input. This preserves temporal features and removes the need for direct coordinate transformation or transformation with motion compensation\footnote{Radar point cloud aggregation with motion compensation means that several point clouds are transferred to a reference point cloud and their positions in the reference cloud are compensated for the vehicle ego-motion. In other words, motion compensation requires knowledge of the vehicle's ego-motion, while direct aggregation does not.}, making it robust in highly dynamic scenes. \par

\item \textbf{Lightweight:} The proposed method uses simple yet effective neural network backbones for feature extraction, where the multi-layer perceptron (MLP) is used for spatial features and the recurrent neural network (RNN) is used for temporal features. The resulting model is lightweight (0.15 M parameters) while providing critical information for understanding vehicle motion and other downstream perception tasks. \par

\item \textbf{Dataset:} As no existing radar dataset fully supports the objective of the proposed method, this work reorganized the ground-truth (GT) class labels of the RadarScenes dataset \cite{radar_scenes_dataset}. Specifically, vehicle ego-motion was used to separate static from non-static objects; the output of the \textit{DeepEgo+} approach was incorporated to compensate for the effects of vehicle acceleration \cite{zhu2025deepego+}, which can otherwise cause mislabeling of static objects; and moving versus non-moving objects were subsequently classified using the dataset’s original labels.

\end{enumerate}

Finally, it must be noted that the goal of this study is different from previous studies on radar-based segmentation. While previous studies have focused on assigning detailed class labels, which is undoubtedly important and meaningful, this study began by seeking a continuation of traditional segmentation, but ended by filling an important gap in the radar perception processing chain. Therefore, it is unfair to compare this work via previous research aiming only at assigning class labels to pixels or points. Instead, the proposed method should be viewed as complementary to traditional radar segmentation and to other radar perception tasks. \par 

The rest of this paper follows this structure. Section \ref{related_work} provides an overview of existing research on this topic. Section \ref{methodology} presents the detailed design of the proposed method. Section \ref{results} first introduces the testing radar dataset and evaluation metrics, and then measures the performance of the proposed method. Finally, Section \ref{conclusions} draws conclusions and outlines future research directions.

\section{Related Works}\label{related_work}
This section reviews the relevant literature. It first outlines prior work on radar-based segmentation and recent advances in the field. It then examines studies on other radar perception tasks to highlight the importance of performing the proposed dual tasks on radar point clouds. Finally, a brief summary of the literature review is provided.\par

\subsection{Radar-based Segmentation}\label{related_work_1}
According to its objectives, previous radar-based segmentation studies can be divided into three categories: semantic segmentation \cite{schumann2018semantic,schumann2019scene,ouaknine2021multi,zeller2022gaussian,fent2023radargnn,zhang2023spatial,wu2024mask,zhang2024tarss}, which assigns a class label to each radar point (detection); instance segmentation \cite{liu2022deep,xiong2022contrastive,zeller2023radar}, which not only classifies each point but also distinguishes between individual objects within the same class; and panoptic segmentation \cite{zeller2024semrafiner}, which combines both approaches by providing semantic labels for all points while also separating instances for object classes. In addition, previous studies can also be divided according to the format of radar data, where except \cite{wu2024mask,zhang2024tarss,ouaknine2021multi}, which use radar cubes (before detection), all of the rest use radar point clouds (after detection). Methods using radar cubes claim they are superior in segmenting small objects, as information can be lost during the detection process \cite{ouaknine2021multi,wu2024mask}. Nevertheless, there are currently no conclusive experimental comparisons demonstrating their effectiveness. For methods that rely on radar point clouds, the RadarScenes dataset \cite{radar_scenes_dataset} appears to be a popular choice since all methods use it to evaluate their performance. Although the RadarScenes dataset provides 10 different classes for moving objects, almost all studies use less than half of them, reflecting the challenges of performing detailed semantic segmentation using sparse and noisy radar data. To handle this challenge, PointNet++ \cite{schumann2018semantic,schumann2019scene,liu2022deep,xiong2022contrastive} and Transformer \cite{zeller2022gaussian,zhang2023spatial,zeller2023radar,zeller2024semrafiner} become the most commonly used feature extraction backbones in these studies. Theoretically, Transformer outperforms PointNet++ in handling sparsity and long-range dependencies; experimentally, Transformer also demonstrates better performance than PointNet++ \cite{zeller2024semrafiner,zhang2023spatial,zeller2022gaussian}.

Based on this brief literature review, it is evident that many studies have extensively explored the topic of radar-based segmentation from various perspectives, such as in terms of objectives, data formats, and feature extraction backbones. It is a solid start, especially in such a pioneering field as automotive radar. However, there are still areas for further improvement. Firstly, except for studies using stationary ego-vehicle datasets \cite{zhang2024tarss,ouaknine2021multi}, nearly all prior research requires knowledge of vehicle ego-motion provided by odometry sensors. Ego-motion is often used to compensate for the measured radial velocity, which is then used as an important input object feature. However, if ego-motion is known, segmenting the static background becomes straightforward, as was done in \cite{schumann2019scene}. Furthermore, since almost 97\% of radar detections come from static objects \cite{schumann2018semantic}, the computational complexity of these methods can be significantly reduced by removing static points from the input. Also, with the compensated radial velocity, it is understandable that most studies achieve scores exceeding 99\% on the Intersection over Union (IoU) metric for classifying `static'\footnote{In the RadarScenes dataset, radar detections from static objects and false positives are both labeled as `static', whereas in this study, they are treated separately.} objects. Last but not least, relying on external sensors may compromise sensor independence and system robustness due to potential erroneous outputs or synchronization issues. \par

Secondly, to address the sparsity problem, some studies \cite{schumann2018semantic,schumann2019scene,zhang2023spatial,fent2023radargnn} rely on combining radar point clouds over a fixed time period (e.g., 500 ms), regardless of the number of clouds aggregated. However, this approach can adversely increase inference latency and system memory consumption \cite{zeller2024semrafiner}. In contrast, other studies \cite{zeller2024semrafiner,zeller2022gaussian,zeller2023radar,zeller2025radar} also merge clouds, but they only allow each radar to contribute once per fused cloud. Given a 60 ms update rate per radar, this can shorten aggregation time while still benefiting from the increased cloud density due to overlapping fields of view (FoVs). Nevertheless, all of the above solutions still introduce some degree of inference latency. Moreover, without motion compensation, they may experience performance degradation in highly dynamic scenes, especially when moving objects are present in the overlap region. Furthermore, since the radars in the RadarScenes dataset are fully unsynchronized\footnote{The time intervals between individual radar outputs are not uniform, and radar transmit and receive operations are not ordered.}, fusing radar point clouds cannot be done directly but requires a heuristic process \cite{zeller_2023_10203864}.  \par

Thirdly, to handle the challenging task of labeling objects in sparse and noisy radar point clouds, previous studies usually adopt NNs with sophisticated feature extraction backbones, such as Transformer and PointNet++. While Transformer outperforms PointNet++, they are typically too bulky to be suitable for radar processing systems that require real-time prediction and immediate feedback \cite{liu2022deep}. In addition, these backbones are often described as `data-hungry', but large radar datasets are expensive to generate and annotate. In any way, due to the fundamental limitations of radar sensors, the performance gains from using complex backbones are not as significant as with optical sensors \cite{reichert2025real,zhang2024semantic}, leading one to wonder: why not use radar for tasks that are better suited to its characteristics? For example, recent studies \cite{zeller2025radar,pan2024ratrack} no longer search for specific object types or bounding boxes, but instead focus on a simpler task of class-agnostic segmentation and tracking. \par

Last but not least, it is worth noting that most previous studies have only focused on segmenting moving objects from radar point clouds, labeling static objects and false positives together as `static'. From the perspective of various radar perception tasks, it is important to conduct a comprehensive segmentation, the reasons for which will be further explained in the next section. \par

\subsection{Other Radar-based Tasks}\label{related_fusion}
A radar point cloud typically contains a mix of detections from moving objects (e.g., vehicles), static objects (e.g., buildings), and false positives (e.g., false detections from sidelobes). Most existing radar-based segmentation tasks focus on separating moving objects, leaving static objects mixed with false positives. However, static objects also play a vital role in many radar perception tasks. For example, the measured radial velocity of static objects can be used to estimate the vehicle's ego-motion \cite{kellner2013instantaneous,kellner2014instantaneous} and thus calibrate the radar's extrinsic parameters \cite{grebner2023self,bao2020motion}. Additionally, knowing where static objects are located allows for the implementation of algorithms such as semantic grid mapping \cite{schumann2019scene}, simultaneous localization and mapping (SLAM) \cite{holder2019real,li2020millimeter}, and amplitude and phase calibration \cite{petrov2021auto}. Furthermore, separating static points from the radar point cloud can help perform free space detection \cite{li2018high,ronecker2024deep}, road course estimation \cite{giese2017road,xu2020road}, and multi-object tracking \cite{pearce2023multi}. Among these studies, most rely on knowing the vehicle's ego-motion provided by external sensors to localize static objects; some use neural networks \cite{zhu2023deepego,zhu2025deepego+}; and some assume a majority of static points and employ one additional processing step such as RANSAC \cite{kellner2013instantaneous} or M-Estimator Sample Consensus (MSAC) \cite{holder2019real}. However, while these solutions can help localize static objects, they leave moving objects mixed with false positives. \par

\subsection{Summary}\label{related_work_3}
In summary, it is essential to point out that in the current radar perception processing chain, there is a missing component that can not only explicitly but also simultaneously segment static objects, moving objects, and false positives from the radar point cloud, which, according to the literature review, is considered crucial for various downstream applications. Furthermore, as the first processing unit after CFAR detectors, this component should be able to work independently, extract important segmentation features automatically from sparse and noisy radar point clouds, and provide fast, accurate, and reliable predictions. The realization of this component summarizes the goals of this research, which will be further described in the next section. \par

\section{Methodology}\label{methodology}
\begin{figure*}[!t]
\centering
\includegraphics[width=\textwidth]{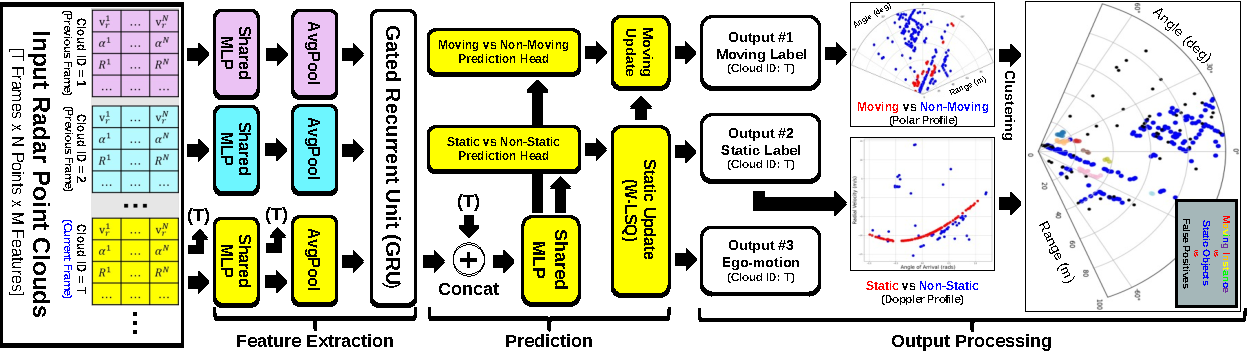}
\caption{Architecture of the proposed neural network for simultaneous static-moving object segmentation and vehicle ego-motion estimation. The network takes multidimensional radar point clouds as input, performs automatic spatial-temporal feature extraction, predicts static labels and moving labels for each detection point, and implements the weighted least squares (w-LSQ) for ego-motion estimation. As an illustrative application, moving instances can be generated after applying a clustering algorithm to the grouped moving objects.}
\label{fig:method}
\end{figure*}

Figure \ref{fig:method} presents the architecture of the proposed method for simultaneous static-moving object segmentation and vehicle ego-motion estimation. The proposed method: takes unprocessed radar point clouds as input, which will be detailed in Section \ref{input}; performs automatic spatial-temporal feature extraction, as explained in Section \ref{feature}; predicts static and moving objects and estimates ego-motion in Section \ref{prediction}; and finally outputs detailed object type labels and moving object instances after several processing steps as detailed in Section \ref{out_processing}. Implementation details will be presented in Section \ref{implementation}. \par

\subsection{Network Input Analysis}\label{input}
\begin{figure*}[!t]
\centering
\includegraphics[width=\textwidth]{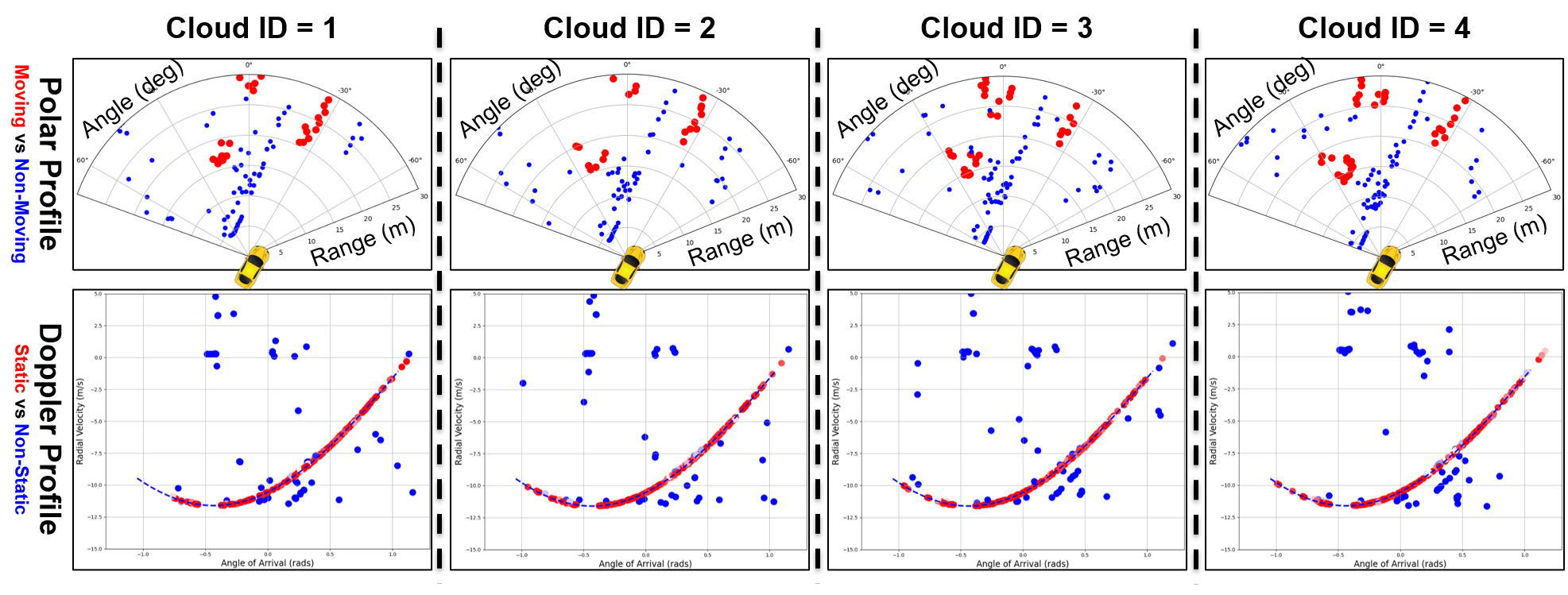}
\caption{An illustration of how moving and static objects appear in radar point clouds across multiple consecutive frames. The first row shows the polar profile, which presents the radar point cloud in the Range-AoA domain. The moving objects, marked in red, exhibit clear spatial concentration and temporal correlation in the polar profile. The second row shows the Doppler profile, which presents the radar point cloud in the radial velocity-AoA domain. The static objects, marked in red, exhibit distinct sine-like spatial pattern with little temporal variation. In this example, the RadarScenes dataset \cite{radar_scenes_dataset} is used.}
\label{fig:method_spatial_temporal}
\end{figure*}

The proposed method takes $T$, unprocessed and chronologically ordered, radar point clouds as input. Typically, the radar point cloud is generated after the application of CFAR algorithms. Each point cloud is assumed to have $N$ radar detection points, and each detection point contains $M$ object features. In this work, $M$ is assumed to be greater than or equal to 3, thus containing at least the uncompensated radial velocity ($v_r$), range ($R$), and angle of arrival (AoA) ($\alpha$) information of the detected objects. The reason for including at least the three selected object features and having $T$ consecutive radar clouds is that they contain the necessary spatial and temporal features for the network to distinguish between moving and static objects, which can also be visually seen in Figure \ref{fig:method_spatial_temporal}. \par

In the Doppler profile, not only is there a clear spatial distinction between static and non-static objects, but there is also a strong temporal correlation between consecutive point clouds of the static detections. In this work, static objects refer to detection points whose measured radial velocity is solely determined by the measurement angle and ego-radar motion, thereby forming a characteristic sine-like pattern in the Doppler profile. In contrast, non-static objects are detection points that deviate from this pattern due to additional velocity contributions, such as independent target motion or false positives. The temporal correlation of static detections is dominated by the continuous motion of the ego-vehicle; consequently, the sine-like pattern remains stable over time, enabling estimation of the radar/vehicle motion from the measured features of static objects, see e.g., \cite{kellner2013instantaneous}. \par

In the polar profile, there are also spatial and temporal correlations between objects in consecutive point clouds. However, because the ego-vehicle is moving and there is no motion compensation, all objects appear to `move' across frames. Furthermore, due to the nature of radar data, the shape and density of detected objects may vary between frames, making reliable discrimination of static objects more challenging. In contrast, detection points from moving objects are usually more spatially concentrated in the polar profile than in the Doppler profile, especially when they are near the radar. This is because, in the Doppler profile, the measured radial velocity at different points on a moving object can vary greatly depending on the measurement angle. Thus, once static objects are first separated in the Doppler profile, the polar profile can help refine the identification of moving objects. In this study, moving objects are defined as detection points originating from targets physically in motion at the time of measurement, whereas non-moving objects comprise static detections, false positives, and inherently mobile targets that are currently stationary (e.g., parked or waiting vehicles).

In summary, unlike previous studies, the proposed method does not require point cloud aggregation, knowledge of the vehicle's ego-motion, or compensation for radial velocity or ego-motion. In contrast, the authors believe that using $T$ consecutive raw radar point clouds is sufficient to simultaneously distinguish between static and moving objects and estimate the vehicle's ego-motion. Regarding the latency issue, for real-time applications, the requirement of $T$ radar frames can be formulated as a moving window so that the proposed method can provide instantaneous predictions. Lastly, the remaining issues are how to effectively extract relevant features for segmentation, which will be detailed in the next section.\par

\subsection{Feature Extraction}\label{feature}
As a result of clear spatial distinctions and strong temporal correlation in the input radar point clouds, the proposed method is able to perform effective feature extraction with simple neural network backbones. For spatial feature extraction, this work employs the PointNet architecture \cite{qi2017pointnet}, which consists of a shared multi-layer perceptron (MLP) followed by an average pooling. Specifically, the MLP is applied independently to each radar detection point in each input point cloud. Afterwards, the pooling layer is used to aggregate a global feature vector for each input point cloud. Despite its simple architecture, the combination of MLP and average pooling has demonstrated effectiveness in extracting the sine-like spatial feature for static object segmentation and vehicle ego-motion estimation \cite{zhu2023deepego,zhu2025deepego+}. Furthermore, because the MLP is shared across input point clouds, the network complexity does not increase with the number of input point clouds. However, it must be acknowledged that this combination has limited ability to capture relationships between neighboring detection points and may therefore be insufficient for tasks requiring fine-grained spatial understanding. Nevertheless, given the sparse radar point clouds, it remains to be seen how much performance improvement more advanced feature extraction backbones (with local details) can bring, as a previous exploration has shown only modest gains \cite{zhu2024hierarchical}. \par

For temporal feature extraction, the proposed method uses the gated recurrent unit (GRU). GRU is a type of recurrent neural network (RNN) that can extract long-term dependencies in sequential data. In this study, the global feature vectors generated by the previous pooling layer are first arranged in chronological order. The GRU then processes these feature vectors sequentially, capturing the hidden relationships within them and outputting a feature vector that contains both spatial and temporal information. It is important to mention that the temporal dependencies between radar point clouds are governed by the continuous motion of the ego-vehicle and moving objects. However, since the input data has no radial velocity compensation or motion compensation, the authors hypothesize that temporal feature extraction is more beneficial for the segmentation of moving objects, while static objects already provide strong differentiation in spatial features, and temporal features are only supplementary. \par

\subsection{Prediction}\label{prediction}
The previous section extracts spatial features from the input radar point cloud and captures the temporal dependencies caused by the continuous object motion. This section explains how to make predictions for each radar detection point. Firstly, the generated spatial-temporal feature vector by the GRU is backpropagated to the original input point cloud and the outputs of different layers in the first shared-MLP through feature concatenation. The concatenation outputs a 2D matrix that still contains $N$ points in one dimension, but in the other dimension contains more global and spatial details in addition to the original $M$ input features. Then, another shared-MLP acts as a decoder, refining the fused features and producing a rich feature vector (per-point) that is based on both spatial-temporal context and local details. After that, the decoder output is sent to two prediction heads, one for static and non-static prediction (static head) and the other for moving and non-moving prediction (moving head). Each head consists of three 1D convolutional layers with the last layer having a sigmoid activation function. The static head outputs a $N \times 1$ vector, where each element contains a value from 0 to 1, indicating the probability of being non-static (0) or static (1). The moving head functions similarly, with its elements representing the probability of a detection point being non-moving (0) or moving (1). \par

Until here, the output of the static head is sufficient for the task of ego-motion estimation. However, since one of the goals is to localize all static objects, the chosen feature extraction backbone has limited ability to capture local context, which is the price of a lightweight network. Consequently, some static objects may be misclassified as non-static and assigned lower weights in the static head, or misclassified as moving and assigned higher weights in the motion head. To address this issue, the proposed method employs two update heads: one for the static weight update and the other for the moving weight update. The initial prediction of the static weight is updated first, based on the fact that knowing the radar motion helps to localize all static objects. Therefore, in the static update head, initial static weights are used to first compute the radar motion via the weighted least squares (w-LSQ) method. Then the estimated radar motion is used to update the static weights for all detection points, as formulated below: \par

\begin{equation}\label{eq:1}
V_{radar} = (A \times W_{static}^{ini} \times A)^{-1}A^T \times W_{static}^{ini} \times D
\end{equation}

\begin{equation}\label{eq:2}
W_{static}^{new} = \frac{1}{\sigma\sqrt{2\pi}} \times \text{exp}(-\frac{\left(A \times V_{radar}-D\right)^2}{2 \times \sigma^2})
\end{equation}

\begin{equation}\label{eq:3}
D = \left[\begin{matrix}-v_{r}^1\\\dots\\-v_{r}^N\end{matrix}\right],\ A = \left[\begin{matrix}cos(\alpha^1)&sin(\alpha^1)\\\dots&\dots\\cos(\alpha^N)&sin(\alpha^N)\\\end{matrix}\right]
\end{equation}

\begin{equation}\label{eq:4}
W_{static}^{ini} = \left[\begin{matrix}w_{static}^{ini,\ 1}&0&0\\0&\cdots&0\\0&0&w_{static}^{ini,\ N}\\\end{matrix}\right],\ V_{radar} = \left[\begin{matrix}v_{x}\\v_{y}\end{matrix}\right]
\end{equation}

Where $v_r$ is the measured radial velocity, $\alpha$ is the AoA measurement, $\sigma$ is the standard deviation of the assumed Gaussian error distribution in the radial velocity measurement, $W_{static}^{ini}$ is the diagonal matrix which contains the predicted initial static weights, $W_{static}^{new}$ is the vector of updated static weights, and $V_{radar}$ is the estimated radar velocity on its x- and y-axes. As for updating the moving weights, the method uses the assumption that a detection point cannot have high weights in both the static head and the moving head, which means that an object cannot be stationary and moving at the same time. Therefore, the updated static weights are used to refine the initial moving weights, as shown below: \par

\begin{equation}\label{eq:5}
w_{mov}^{new,\ n} =
\begin{cases} 
w_{mov}^{ini,\ n} & w_{static}^{new,\ n} \leq c_{static} \\ 
0 & w_{static}^{new,\ n} > c_{static}
\end{cases}
\end{equation}

\begin{figure*}[!t]
\centering
\includegraphics[width=\textwidth]{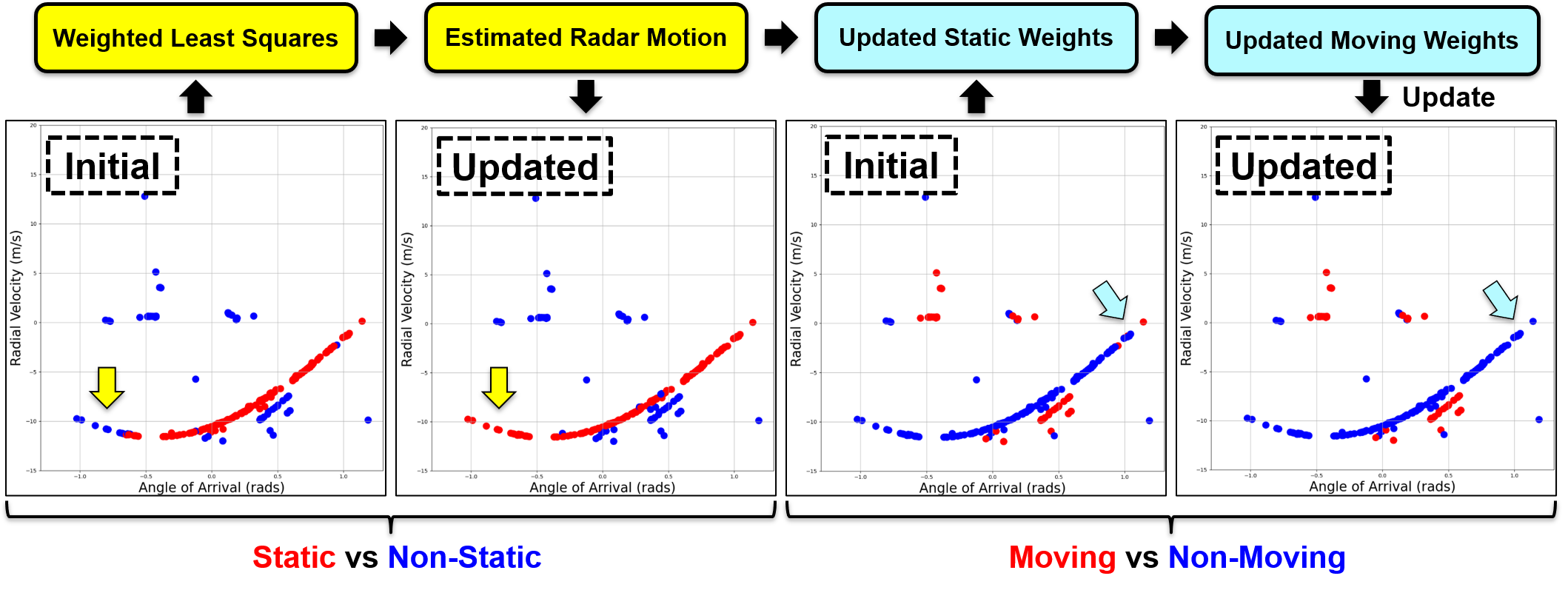}
\caption{An illustration of how the initial weights for static and moving objects are updated in the two weight update heads. The yellow blocks represent the static update head, and the cyan blocks represent the moving update head. In this example, the RadarScenes dataset \cite{radar_scenes_dataset} is used, and the plots show the radar point cloud in the radial velocity-AoA domain.}
\label{fig:weight_update}
\end{figure*}

Where $c_{static}$ is the empirical parameter of the threshold. Lastly, Figure \ref{fig:weight_update} presents a visual illustration of the update process of the static weights and the moving weights. \par

Finally, it is important to clarify that although a single 3-class prediction head (moving–static–false positives) is possible, the problem exhibits a hierarchical structure, as shown previously. In this hierarchy, the initial static prediction provides the basis for estimating radar motion, which in turn updates the initial static prediction and cross-checks subsequent moving predictions. Using two prediction heads allows the architecture to explicitly encode this structure, enabling the network to solve simpler binary classification tasks rather than implicitly learning the full set of relationships. Moreover, the two-head approach avoids the need to directly model false positives, which is inherently ill-defined and highly variable, thereby improving robustness and overall classification accuracy. \par

\subsection{Output Processing}\label{out_processing}
For the current radar input point cloud, the proposed method can simultaneously estimate the vehicle's ego-motion and provide labels for static and moving objects. Given the radar extrinsic parameters and the estimated radar motion, the vehicle motion can be computed as follows:  \par

\begin{equation}\label{eq:2_2}
V_{car} = \left[\begin{matrix}v_{x}^{car}\\\omega\end{matrix}\right]=\left[\begin{matrix}v_{x}\cdot cos(\theta)-v_{y}\cdot sin(\theta)+y\cdot\omega\\\frac{1}{x} \cdot (v_{y} \cdot cos(\theta)+v_{x} \cdot sin(\theta))\end{matrix}\right]
\end{equation}

Where $v_{x}^{car}$ is the vehicle's translational speed, $\omega$ is its rotation rate, and the vehicle is assumed to have no lateral speed, i.e., $v_{y}^{car}=0$. $x$, $y$, and $\theta$ are the mounting position and angle of the radar sensor with respect to the rear center of the vehicle. The labels for static and moving objects can be obtained directly by applying thresholds to the updated static and moving weights respectively. In this study, both thresholds are set empirically to 0.1. As shown in the rightmost sub-figure of Figure \ref{fig:method}, the static and moving labels can be merged together to achieve a clear separation of false positives. Furthermore, since moving objects are explicitly separated, clustering algorithms such as the DBSCAN can be applied to them to achieve moving instance segmentation. However, the instance segmentation is just one illustrative example, and as discussed in Section \ref{related_work}, many radar perception tasks can be connected to the output of the proposed method. \par

\subsection{Implementation Details}\label{implementation}
The proposed method is trained with one Nvidia A100 GPU provided by the Delft High Performance Computing Centre (DHPC) \cite{DHPC2024}. The batch size is 64 and the maximum training epoch is 400, but training can be stopped when the training loss stops improving after more than 10 epochs. The Adam optimizer is used and the initial learning rate is 0.001. The learning rate is decreased by a factor of 0.5 when the training loss stops improving after more than 5 epochs. The tuning parameter $\sigma$ is empirically set to 0.013. For the shared-MLP, it contains three 1D convolutional layers, each followed by a batch normalization layer and a ReLU layer for non-linearity. In the second shared-MLP (the decoder), the second 1D convolutional layer is followed by an additional dropout layer with a dropout rate of 0.3, and the randomly generated dropout mask is identical for the feature vector of each detection point. Finally, this study uses two cross-entropy losses to measure the difference between the predicted results and the true values of static labels and moving labels, respectively. Since the loss of ego-motion estimation is closely related to the loss of static prediction, errors in ego-motion are not backpropagated. To mitigate the influence of low-quality training examples, the final loss is the sum of the two cross-entropy losses multiplied by the sample weight (described in more details in \cite{zhu2025deepego+}).

\section{Results and Discussion}\label{results}
This section presents the evaluation results of the proposed method. Specifically, the used radar dataset and the generation of ground truth will be introduced first, followed by a comprehensive performance study of the proposed method and related methods in the literature. \par

\subsection{Radar Dataset}\label{results:dataset}
Following the practice in previous studies on ego-motion estimation \cite{zhu2023deepego,zhu2025deepego+} and radar segmentation \cite{zhang2023spatial,zeller2024semrafiner}, this study uses the RadarScenes dataset \cite{radar_scenes_dataset} to evaluate the proposed method. RadarScenes is a challenging radar dataset collected from real-world traffic and driving. During data collection, four automotive radars were installed on the front of the vehicle, two of which faced forward (hereinafter referred to as `Radar 2' and `Radar 3'), and the other two faced the side (hereinafter referred to as `Radar 1' and `Radar 4'). After collection, the radar data is preprocessed to generate radar point clouds containing detection range, AoA, radial velocity, and radar cross section (RCS). In addition, moving objects are manually annotated by human experts and classified into 10 different object categories. The ego-motion information of the vehicle is recorded using the vehicle’s odometry sensors and a differential global positioning system (DGPS). \par

Although the RadarScenes dataset records accurate vehicle motion and provides manually labeled point clouds, due to the new task proposed in this study, four additional processing steps are required, in order to generate ground truth (GT) data for model training and evaluation. Firstly, radar detections of static objects and false positives are not individually labeled in the dataset. To distinguish them, the recorded vehicle motion is used to localize static detections from the radar point cloud. Specifically, similar to Eq. \ref{eq:2}, the vehicle motion is first transformed to radar motion, and then the GT static labels can be calculated. For moving objects, the GT moving labels are generated based on the class labels provided by the dataset, where 0 represents non-moving (`Class 11') and 1 represents moving (`Class 1 to 10'). If a detection is neither labeled as static nor moving, it is defined as one of the false positives. Conversely, if a detection is labeled as both static and moving due to, for example, mislabeling in the GT, it is corrected to static but non-moving, as vehicle GT motion is more reliable and trustworthy than human annotations. \par

In the second processing step, the effect of vehicle acceleration on the measured radial velocity is resolved. As detailed in \cite{zhu2025deepego+} Section-IV-G (Fig. 9-(c)), due to vehicle non-zero acceleration, the Doppler frequency and the associated phase shift will vary with slow time and the estimated radial velocity will not match the vehicle velocity. Therefore, the GT static labels generated solely based on vehicle motion may be inaccurate. As shown in \cite{zhu2025deepego+}, \textit{DeepEgo+} can mitigate this effect by using a two-step signal processing with NNs. The first step locates the static detection points, and the second step compensates for the effect of non-zero acceleration and estimates the vehicle ego-motion. Therefore, in this study, if the \textit{DeepEgo+} ego-motion estimation error is below a preset threshold, its output is used to help better localize static objects; otherwise, the vehicle's GT motion is used.

Thirdly, the RadarScenes dataset contains 158 2-minute-long individual sequences from each of the four radars. In almost half of the sequences, the ego-vehicle is (or almost is) stationary and monitors moving objects. This is well-suited for tasks such as object detection and motion segmentation. However, for the dual task of ego-motion estimation and static-moving object segmentation, a dataset containing an ego-vehicle in constant motion is desired for both training and evaluation. Therefore, this work uses a minimum driving distance of 500 meters for sequence selection, resulting in 63 radar sequences captured in challenging scenarios such as highways and city traffic. Nevertheless, these 63 radar sequences still contributed more than 2 hours of recording time, which is equivalent to a driving distance of more than 70 km. It is worth mentioning that the reduction in the number of sequences necessarily increases the difficulty of the task, because not only is it more difficult and meaningful to distinguish between static and moving objects when the vehicle is in constant motion rather than stationary, but also training neural networks on smaller datasets can lead to some well-known challenges such as overfitting and generalization problems.\par

The last processing step is to deal with short-lived labeled moving objects. As shown in Table \ref{tab:lifespan}, due to different installation angles, the number of moving objects observed by the four radars varies greatly. Radar 1 faces the side of the street and picks up minimal objects, while Radar 2 and Radar 3 face forward, cover both lanes, and pick up the most objects. In addition, since Radar 1 and Radar 4 face sideways, moving objects often appear at close range and enter and leave the radar field of view quickly, resulting in a very short lifespan. Moving objects with short lifespans contain little temporal features and may confuse model training and increase false alarm rates. Therefore, moving objects with a lifespan shorter than 5 radar frames (around 0.3 s) are labeled as non-moving for the sake of training. Also, the data from Radar 1 is not used for performance evaluation because there are only about 10 moving objects per sequence on average. Finally, unless otherwise specified, the following experiments are all conducted using Radar 3 data, and the `leave-one-out'\footnote{The test radar sequences are taken out, one by one, from the selected 63 radar sequences, and the remaining sequences are used for model training and validation following the 80\%-20\% rule. After all 63 sequences have been used once as the test sequence, the final performance of the tested method is measured and averaged.} training, validation, and testing strategy is adopted so the performance on `unseen' data can be measured.\par

\begin{table*}[!t]
    \caption{The radar mounting position and the number of labeled moving objects in the selected 63 radar sequences from the RadarScenes \cite{radar_scenes_dataset} dataset.}
    \centering
    \begin{tabular}{|c|c|c|c|c|}
    \hline
    \textbf{Radar Name} & \textbf{Radar 1} & \textbf{Radar 2} & \textbf{Radar 3} & \textbf{Radar 4} \\ \hline
    \textbf{Pointing Direction} & Side-looking & Front-facing & Front-facing & Side-looking \\ \hline
    \textbf{Labeled Moving Objects} & 756 & 3484 & 3685 & 2396\\ \hline
    \textbf{Object's Lifespan $<$ 5 Frames} & 106 & 407 & 207 & 227 \\ \hline
    \end{tabular}
    \label{tab:lifespan}
\end{table*}

\subsection{Evaluation Metrics}
Due to the dual task of the proposed method, this study proposes a series of evaluation metrics to measure its performance in ego-motion estimation and static-moving object segmentation. For moving object segmentation, inspired by one downstream application of radar-based object tracking \cite{tan2023tracking}, the moving objects predicted by the proposed method and identified by the GT labels are first clustered into moving instances, respectively. The density-based spatial clustering of applications with noise method (DBSCAN) \cite{ester1996density} is used for clustering. Then, the grouped moving objects are converted into point target lists by finding the average position of all points in the same cluster. Afterwards, the cost matrix is calculated based on the L2 distance between the point objects in the GT list and the prediction list. Next, the Jonker–Volgenant algorithm \cite{jonker1986improving} is implemented to solve the data association problem. Based on its output, three numbers can be determined for a given radar frame: the number of correctly detected moving objects (TP), the number of false detections (FP), and the number of missed detections (FN). Finally, the TP, FP, and FN of all radar frames are summed separately, and the following evaluation metrics can be calculated: \par

\begin{enumerate}
    \item \textbf{False Discovery Rate (FDR)} shows the proportion of false detections among all detected moving instances. In other words, it reflects the frequency of false detections. FDR is defined as follows: \par

    \begin{equation}
        FDR = \frac{FP}{FP + TP}
    \end{equation}
    
    \item \textbf{Missed Detection Rate (MDR)} measures how often true moving instances are misclassified as non-moving. It is defined as follows: \par

    \begin{equation}
        MDR = \frac{FN}{FN + TP}
    \end{equation}
    
    \item \textbf{F1 Score (F1)} is the harmonic mean of Precision and Recall. Therefore, the F1 Score will be high only when both Precision and Recall are high. This property makes it well-suited for summarizing detection performance, especially in the case of class imbalance. It is defined as follows: \par

    \begin{equation}
        F1 = \frac{2 * TP}{2 * TP + FP + FN}
    \end{equation}

    \item \textbf{Intersection over Union (IoU)} is a commonly used evaluation metric for computer vision tasks such as detection and segmentation. Traditionally, it is computed geometrically based on the overlap between the predicted region (for example, the bounding box) and the actual region. However, due to the characteristics of radar sensors, the shape of detected objects changes with distance and angle, and they have fewer geometric features due to low azimuth resolution. In addition, the actual area may also be erroneous and incomplete due to errors in the GT label. Therefore, in order to adapt to the radar characteristics, this work defines the IoU metric as follows:\par

    \begin{equation}
        IoU = \frac{TP}{TP + FP + FN}
    \end{equation}

    Here TP represents the correct overlap, FP represents the extra predicted moving instances, and FN represents the missed GT instances. \par
    
\end{enumerate}

Since the performance of static object segmentation is closely related to the performance of vehicle ego-motion estimation, static segmentation is not explicitly evaluated in this study. Therefore, only the motion error of the tested method is reported, and the following two metrics proposed in \cite{zhu2023deepego,zhu2025deepego+} are used: \par

\begin{enumerate}
    \item \textbf{Saturated Root Mean Square Error (S-RMSE)} is a truncated version of RMSE. It measures estimation accuracy like RMSE, but is less sensitive to `outliers' than RMSE. For example, when using RMSE, ego-motion estimation performance can be significantly biased by large errors in the GT. To handle it, following the definition in \cite{zhu2025deepego+}, S-RMSE can be expressed as follows:\par

    \begin{equation}\label{eq:s_rmse_1}
    S\_RMSE(\mathbf{X}_{car}, \hat{\mathbf{X}}_{car}) = \sqrt{\frac{1}{P}\sum_{p=1}^{P}{d^2_p}}
    \end{equation}

    Where:
    
    \begin{equation}\label{eq:s_rmse_2}
    d_p =
    \begin{cases} 
    x^p - \hat{x}^p & | x^p - \hat{x}^p| \leq c_{err} \\ 
    s & | x^p - \hat{x}^p | > c_{err}
    \end{cases}
    \end{equation}
    
    $x^p$ and $\hat{x}^p$ are the ground truth and estimated ego-motion ($\hat{v}_x^{car}$ or $\hat{\omega}$) at timestamp $p$, and $P$ is the total number of timestamps of the tested radar sequence. $c_{err}$ is the predefined range of considered errors, and $s$ is the fixed error assigned when the error exceeds the predefined range. In this work, $c_{err}$ and $s$ are set to 50 ($cm/s$) for measuring errors in $\hat{v}_x^{car}$, and 2.86 ($deg/s$) for measuring errors in $\hat{\omega}$.
    
    \item \textbf{Relative Trajectory Error (RTE)} measures the distance between the end point of the estimated trajectory and the end point of the ground truth trajectory. Since errors accumulate, RTE can reflect the long-term stability of the test method. However, RTE can be sensitive to errors that occur at the beginning, especially when the trajectory is long. In this study, the 63 trajectories of the ego-vehicle are divided into 50-meter segments and the RTE (RTE\_50) is calculated.\par
\end{enumerate}

In summary, the proposed evaluation metrics enable a comprehensive understanding of the performance of the proposed method. For moving object segmentation, different from previous studies, the proposed metric is applied to clustered object lists, which is more suitable for the characteristics of radar data and less sensitive to errors in the GT label. For ego-motion estimation, popular evaluation metrics are taken from the literature. These metrics can not only indicate the accuracy and long-term stability of the tested method in ego-motion estimation, but also indirectly reflect the performance in static object segmentation. Lastly, in addition to these quantitative evaluation metrics, qualitative results are also presented for better visual understanding\footnote{Also presented on \textcolor{blue}{https://www.youtube.com/@RadarTechTUDelft/videos}}. \par

\subsection{Comparisons with State Of The Art (SOTA)}
Before presenting the detailed performance evaluation and comparison, it is worth mentioning that the proposed approach differs from previous studies in two aspects. Firstly, this study aims to achieve both ego-motion estimation and static-moving object segmentation simultaneously, which is a first of its kind and also introduces a different evaluation method. Secondly, motivated by many other radar downstream applications that are premised on separating static \cite{grebner2023self, kellner2013instantaneous, xu2020road, ramesh2021landmark} or moving objects \cite{cao2022phd, schumann2019scene}, this study redefines the conventional objectives in radar segmentation and provides a one-step solution for these applications. Therefore, the authors must acknowledge that it becomes challenging and difficult to make a fair comparison of the proposed method with the state-of-the-art methods (SOTA) in the literature given the above differences. \par 

Table \ref{tab:sota} summarizes a list of representative previous studies in the field of radar-based ego-motion estimation and segmentation. For radar-based segmentation, the closest previous study to this work is \cite{zeller2025radar}, which also performs moving object segmentation, while other studies seek accurate and detailed class labels for moving objects. Nevertheless, the authors believe that the proposed method is more competitive than previous studies in the following aspects. Firstly, all listed works require knowledge of the vehicle's ego-motion. In most cases, ego-motion is used to compensate for the measured radial velocity, which has been shown to be a key feature for identifying static and non-static objects \cite{schumann2018semantic,zeller2022gaussian}. However, if ego-motion is known, the input radar point cloud in these studies can be significantly simplified by removing all static objects, making it easier to distinguish moving objects from false positives and saving computational resources. This is because static objects and false positives together contribute almost 97\% of radar detections in the RadarScenes dataset. Furthermore, dependence on external odometry sensors can undermine sensor independence and reduce system robustness, as this introduces risks of erroneous outputs or synchronization problems. In contrast, the proposed method can independently work on unprocessed radar point clouds and does not rely on any external sensors or motion compensation. The special network design enables it to capture relevant features from the point cloud, thereby not only separating moving objects but also localizing static objects and estimating vehicle motion. \par

\begin{table*}[!t]
    \caption{Comparison between the proposed method and representative studies in the literature. For ego-motion estimation (Ego-M.), \textit{DeepEgo} \cite{zhu2023deepego}
    is selected and its performance is measured in RTE\_50 after training with the same radar sequences as the proposed method. For the segmentation task, four previous studies are selected and their reported performances in terms of IoU and F1 scores are shown in the table.}
    \centering
        \begin{tabular}{| c | c | c | c | c | c | c | c |}
        \hline
        \textbf{References} & \textbf{Radar Task} & \textbf{Main Backbone} & \textbf{Odometry Data} & \textbf{Point Cloud Aggregation} & \textbf{Parameters (M)} & \textbf{IoU / F1} & \textbf{RTE\_50} \\
        \hline
        \cite{zhu2023deepego} & Ego-Motion & MLP & Not Required & Not Required & 0.8 & N/A / N/A & 16.00\\
        \hline
        \cite{zeller2025radar} & Segmentation & Transformer & Required & Fuse Multiple Radars & N/A & 0.81 / N/A & N/A \\
        \hline
        \cite{zeller2024semrafiner} & Segmentation & Transformer & Required & Fuse Multiple Radars & 4.5 & N/A / N/A & N/A \\
        \hline
        \cite{zhang2023spatial} & Segmentation & Transformer & Required & Fuse 500 ms Radar Scans & 7.36 & N/A / 0.81 & N/A \\
        \hline
        \cite{zeller2022gaussian} & Segmentation & Transformer & Required & Fuse Multiple Radars & 8.4 & N/A / 0.80 & N/A \\
        \hline
        Proposed & Ego-M. \& Seg. & MLP & Not Required & Not Required & 0.15 & 0.86 / 0.92 & 1.8 \\
        \hline
        \end{tabular}
    \label{tab:sota}
\end{table*}

Secondly, all listed segmentation tasks perform point cloud aggregation across multiple radars or over a period of time. One reason for this is the low angular resolution of radars, while point cloud aggregation helps enrich the geometric features of objects. However, point cloud aggregation requires good sensor synchronization and the knowledge of the relative extrinsic parameters between radars. Furthermore, without motion compensation, the aggregation effect can deteriorate in highly dynamic scenes, where the shape of objects changes with speed. For example, a fast-moving car may look like an elongated truck after aggregation. In addition, temporal information may be lost after aggregation across multiple radar frames. Moreover, the aggregation process inevitably introduces inference delays, which may affect applications that require a real-time fast response. On the contrary, although the proposed method uses multiple single-frame radar point clouds, they are arranged in time sequence, processed independently, and can form a moving window to provide instantaneous predictions for the current time. \par

Thirdly, previous studies typically employ complex feature extraction backbones (such as Transformer \cite{zhao2021point}) to help capture crucial details so that the exact categories of moving objects can be distinguished in sparse and noisy radar point clouds. However, these backbone networks usually require very large datasets for model training, otherwise there may be risks of overfitting and poor generalization ability, while radar data collection is expensive and labeling sparse radar data is very time-consuming. Furthermore, for automotive applications, these `large' networks typically require more computing resources and can incur higher latency, but the performance gain from using complex backbones for radar segmentation is much smaller than for the same task in LiDAR. Therefore, this study breaks this convention and instead separates moving and static objects, which the authors believe is more appropriate and reliable for radar data, more beneficial for other downstream applications, while also helping to build a lighter network. As shown in the Table \ref{tab:sota}, even for the dual task, the proposed method is the lightest of all listed methods and can be trained using less but more challenging data.

For ego-motion estimation, the previous SOTA method \textit{DeepEgo} \cite{zhu2023deepego} is trained using the same dataset and compared with the proposed method. Firstly, both \textit{DeepEgo} and the proposed method can achieve instantaneous ego-motion estimation without the need for point cloud aggregation and odometry data. In addition, both use lightweight backbone networks for feature extraction. Differently, the proposed method achieves superior ego-motion estimation performance compared to \textit{DeepEgo}. This is primarily because the proposed method leverages temporal information from previous radar frames to better localize static objects and estimate vehicle motion in the current frame. \par

Finally, it is worth mentioning that, except for \textit{DeepEgo}, the segmentation performance (i.e., IoU and F1) of the selected works is directly taken from the corresponding references\footnote{The definitions of IoU and F1 may also differ from this paper.}. This is because the work proposed in this paper differs significantly from previous studies, not only in the main objective but also in the requirements of the size of training data, point cloud processing, and external sensor information. Therefore, performance comparison with compromises in re-implementation will be unfair to either the previous studies or the proposed work.\par

\subsection{Performance over Moving Window Lengths}
\begin{figure}[!t]
\centering
\includegraphics[width=0.4\textwidth]{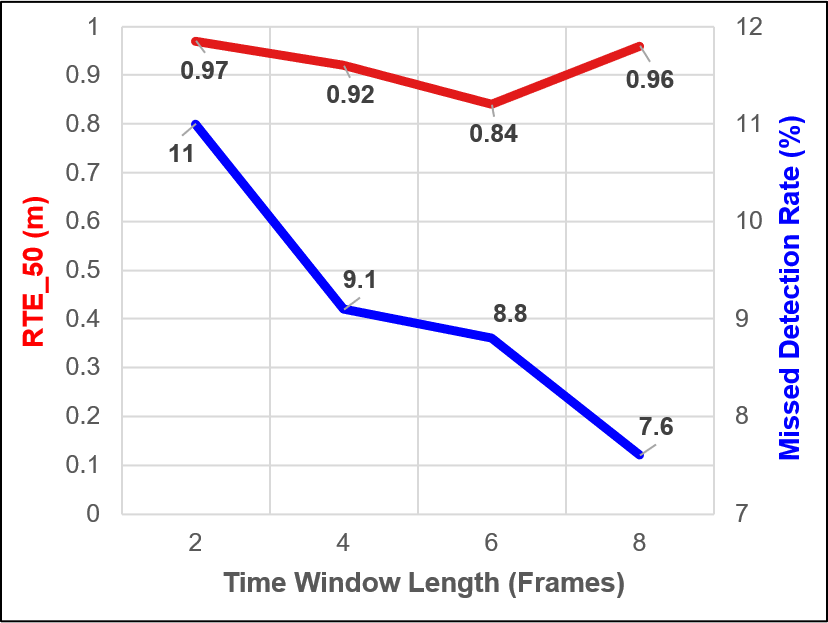}
\caption{The performance of the proposed method for moving object segmentation and ego-motion estimation with different lengths of the input moving window (in radar frames). The blue solid line represents the missed detection rate of the model, and the red solid line represents the RET\_50.}
\label{fig:window_length}
\end{figure}

The length of the input moving window is an important hyperparameter of the proposed method because it determines how many history radar point clouds and how much temporal information the proposed NN can exploit. As explained in Section \ref{feature}, this temporal information is crucial for localizing moving objects since radial velocity is not compensated and the input point cloud is sparse. To show its effect, Figure \ref{fig:window_length} provides the performance of the proposed method for moving object segmentation and ego-motion estimation with different window lengths. As expected, the missed detection rate decreases rapidly with the increase of input length, indicating that more moving objects are correctly segmented.  However, the RTE\_50 metric does not change significantly, reflecting that longer moving window lengths may have little effect on ego-motion estimation performance, which was also expected since static objects already show unique patterns in the single-frame Doppler profile (Figure \ref{fig:method_spatial_temporal}). Finally, unlike point cloud aggregation, moving the window does not affect timely predictions, but it still requires more memory resources than single-frame methods. Therefore, it is recommended to adjust this parameter based on application requirements. In this study, the input window length is set to 8 for all experiments.

\subsection{Performance over Distances}
\begin{figure}[!t]
\centering
\includegraphics[width=0.4\textwidth]{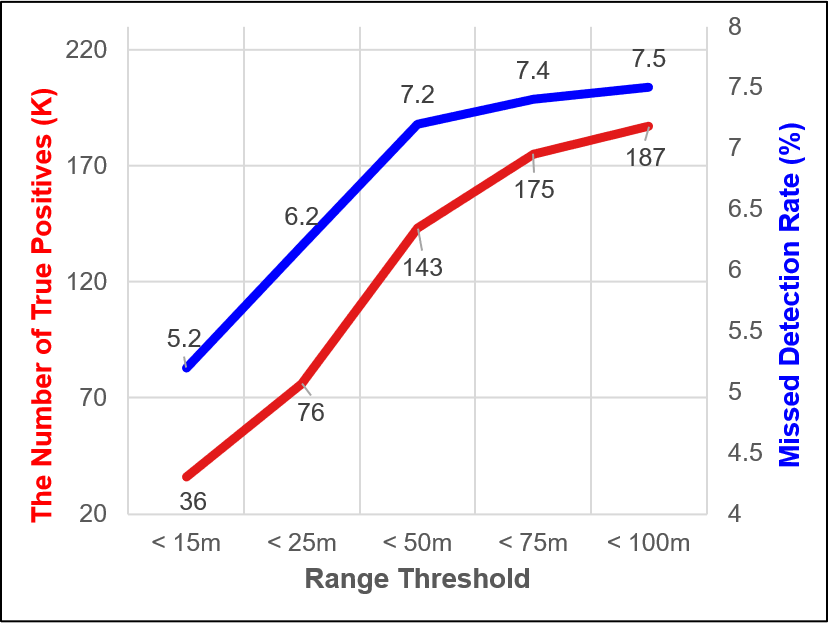}
\caption{The effect of thresholding on the measurement range of Radar 3. The threshold changes the size of the radar field of view, thereby changing the number of moving objects within it. The red solid line shows the relationship between the range threshold and the number of TPs. Note that the number of TPs every moving object can contribute is the same as its lifespan measured in number of frames. The blue solid line shows the relationship between the range threshold and the performance of the proposed method in terms of missed detection rate.}
\label{fig:distance_threshold}
\end{figure}

One of the advantages of radar sensors is their long detection range. The automotive radar used in the RadarScenes dataset can cover a detection range of up to 100 meters. However, the spatial cross-range resolution of radar is finer at a close range and coarser at a long range. This is because, with a fixed azimuth resolution, the area covered by one resolution cell increases with distance, even if the range resolution remains constant. Therefore, distant moving objects may only produce a few detection points, and it is important to understand how this will degrade the segmentation performance of the proposed method. Figure \ref{fig:distance_threshold} shows the missed detection rate of the proposed method measured at different range thresholds. When the radar’s FoV is limited to a maximum range of 15 meters, the proposed method misses only 5.2\% of TPs, e.g., a moving object appears in 100 radar frames but is missed in only about 5 frames. However, as the maximum range increases from 15 to 50 meters, the missed detection rate and the total number of TPs within the radar FoV increase rapidly. Finally, the deterioration slows down after 50 meters, reaching a missed detection rate of 7.5\%, and a total of 187 K TPs are detected within 100 meters. \par

\subsection{Ablation Study on Input Features}
As mentioned in Section \ref{methodology}, the proposed method requires that the input radar point cloud contains at least three types of object features, namely, range, AoA, and radial velocity. To understand which features are important for the task of ego-motion estimation and moving object segmentation, this section conducts an ablation study on the selected input features. As shown in Table \ref{tab:ablation}, first, the radial velocity measurements are the most valuable object feature for both ego-motion estimation and moving object segmentation. For ego-motion estimation, the measured radial velocity and AoA help to clearly distinguish between static and non-static objects, as shown by the Doppler profiles. This can therefore explain the degradation in ego-motion performance when AoA is removed from the input data. For moving object segmentation, even if the angle and range information is preserved, it is very difficult to distinguish between moving and non-moving objects without radial velocity. This is because radial velocity helps separate static objects, making moving objects more visible than the false positives in the radar point clouds. However, it is also interesting to note that even without angle information, the proposed method still retains the ability to detect moving objects, albeit with poor ego-motion estimation performance. Finally, among the three tested input features, the range information appears to have the least impact on the performance of the dual task. This can also be intuitively understood from the Doppler profile, where moving objects, static objects, and false positives also show clear temporal and spatial distinctions across multiple radar frames. However, as predicted in Section \ref{input}, range information becomes more important for nearby moving objects, since these objects can occupy many angular cells and be spatially separated in the Doppler profile. As shown in the table, the performance gap between `All Features' and `No Range' is larger when a 15-meter range threshold is applied. \par

\begin{table}[!t]
    \caption{Effects of different input features on ego-motion estimation and moving object segmentation.}
    \centering
    \begin{tabular}{|c|c|c|}
    \hline
    \textbf{Input Conditions} & \textbf{F1 Score} & \textbf{RTE\_50 [m]} \\ \hline
    \textbf{No Range} & 0.91 & 1.99 \\ \hline
    \textbf{No Azimuth AoA} & 0.85 & 62.3 \\ \hline
    \textbf{No Radial Velocity} & 0.58 & 58.7 \\ \hline
    \textbf{All Features} & 0.93 & 0.96 \\ \hline
    \textbf{No Range ($<15m$)} & 0.93 & N/A \\ \hline
    \textbf{All Features ($<15m$)} & 0.97 & N/A \\ \hline
    \end{tabular}
    \label{tab:ablation}
\end{table}

\subsection{Performance over Radar Positions}
Previous experiments are conducted using data from Radar 3 because this sees the most moving objects, which is in line with the goals of this work. However, it is also important to show that the proposed method can work at other positions or mounting angles. Therefore, in addition to Radar 3, this section also applies the proposed method to data of Radar 2 and Radar 4. As shown in Table \ref{tab:radar_position}, the proposed method performs almost the same on Radar 2 and Radar 3. However, when using data from Radar 4, while the model can still perform good ego-motion estimation, its segmentation performance degrades. One reason for this is that Radar 4 is looking sideways at the passing lane, and moving objects can move perpendicular to the direction the radar is pointing, affecting measured radial velocities. Furthermore, side-looking radars can also capture random objects on the street that are either far away (a few detection points) or briefly within the radar's FoV. To examine scenarios closer to real-world use, a maximum threshold of 15 meters is applied to the detection range of Radar 4, so the radar only covers the overtaking and oncoming lanes. Under this condition, the model performed just as well on Radar 4 as on Radars 2 and 3, demonstrating the effectiveness of the proposed method even under adverse mounting angles.

\begin{table*}[!t]
    \caption{The performance of the proposed method under different radar installation positions and angles. In this experiment, the proposed method is trained and evaluated separately using data from different radars. The last row of the table applies a maximum threshold of 15 meters to the detection range of Radar 4. The model's ego-motion estimation performance at the given threshold is not measured and is therefore marked as `N/A'.}
    \centering
        \begin{tabular}{| c | c | c | c | c | c | c |}
        \hline
        \textbf{Conditions} & \textbf{FDR (\%)} & \textbf{MDR (\%)} & \textbf{F1 Score} & \textbf{S-RMSE Vx (cm/s)} & \textbf{S-RMSE $\omega$ (deg/s)} & \textbf{RTE\_50 (m)}\\
        \hline
        Radar 2 & 6.4 & 7.7 & 0.93 & 0.47 & 0.11 & 1.4 \\
        \hline
        Radar 3 & 6.4 & 7.6 & 0.93 & 0.37 & 0.12 & 0.96 \\
        \hline
        Radar 4 & 11.5 & 16.1 & 0.86 & 2.03 & 0.11 & 0.41 \\
        \hline
        Radar 4 ($<15m$) & 7.8 & 5.0 & 0.94 & N/A & N/A & N/A \\
        \hline
        \end{tabular}
    \label{tab:radar_position}
\end{table*}

\subsection{Qualitative Result: Static-Moving Object Segmentation}
\begin{figure*}[!t]
\centering
\includegraphics[width=\textwidth]{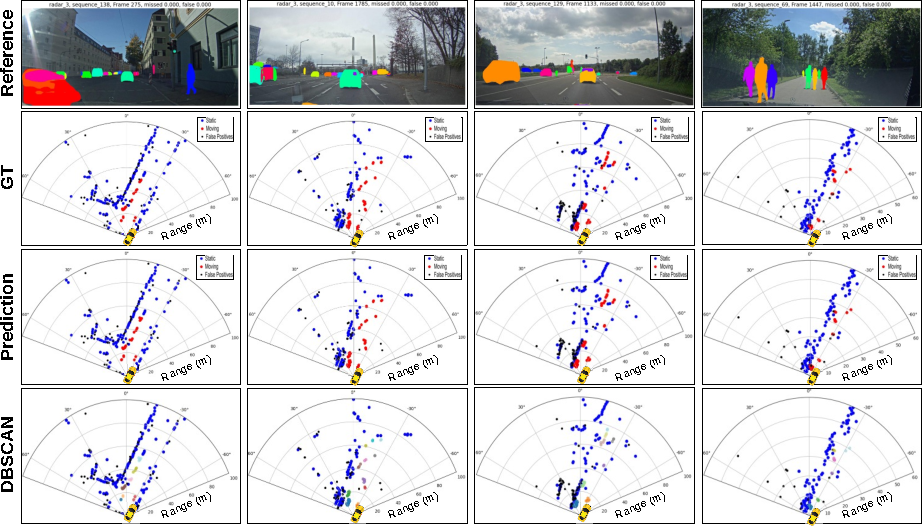}
\caption{Qualitative results of the proposed method for static and moving object segmentation in 2D polar plots. The proposed method is tested in four different driving scenarios (shown in four columns). The first row shows images from an on-vehicle reference camera, the second row shows the ground truth, the third row shows the model's predictions, and the last row shows the clustering output after applying DBSCAN to the predictions. In the second and third rows, moving objects are marked in red, static objects are marked in blue, and false positives are marked in black.}
\label{fig:moving_segmentation}
\end{figure*}

While the previous sections quantitatively evaluated the performance of the proposed method, this section provides qualitative tools for better visual understanding. As shown in Figure \ref{fig:moving_segmentation}, in addition to providing vehicle ego-motion, the proposed method can also achieve simultaneous segmentation of static and moving objects in a variety of challenging scenarios, such as driving on a narrow and busy street, driving at high speed in an open area, or driving but being surrounded by slow-moving pedestrians. Different from previous studies, the proposed method can directly segment sparse radar point clouds without the need for point cloud aggregation. It is also worth noting that the predicted static and moving objects can be used by many radar downstream tasks. For example, as shown in the figure, clustering algorithms such as DBSCAN can be applied to generate moving instances, and then classic multi-target tracking algorithms can be used to estimate their motion states or trajectories. Finally, for detections that are neither labeled as moving nor static, they are classified as false positives in this study. Typically, reflections coming from side-lobes and multipath can be labeled as false positives. However, as shown in the third column of the figure, detections originating from the static treetops to the left of the ego-vehicle are also marked as false positives in both the GT and the prediction. This is because the radar sensors used in the RadarScenes dataset only have azimuth and range resolution, but elevation also affects the measured radial velocity, leading to incorrect predictions and GTs for static objects that are not at the same level as the radar sensor. However, if in future works these detections can be correctly segmented, it may be possible to also estimate their heights \cite{laribi2017new}. \par

\subsection{Qualitative Result: Localization and Mapping}
\begin{figure*}[!t]
\centering
\includegraphics[width=\textwidth]{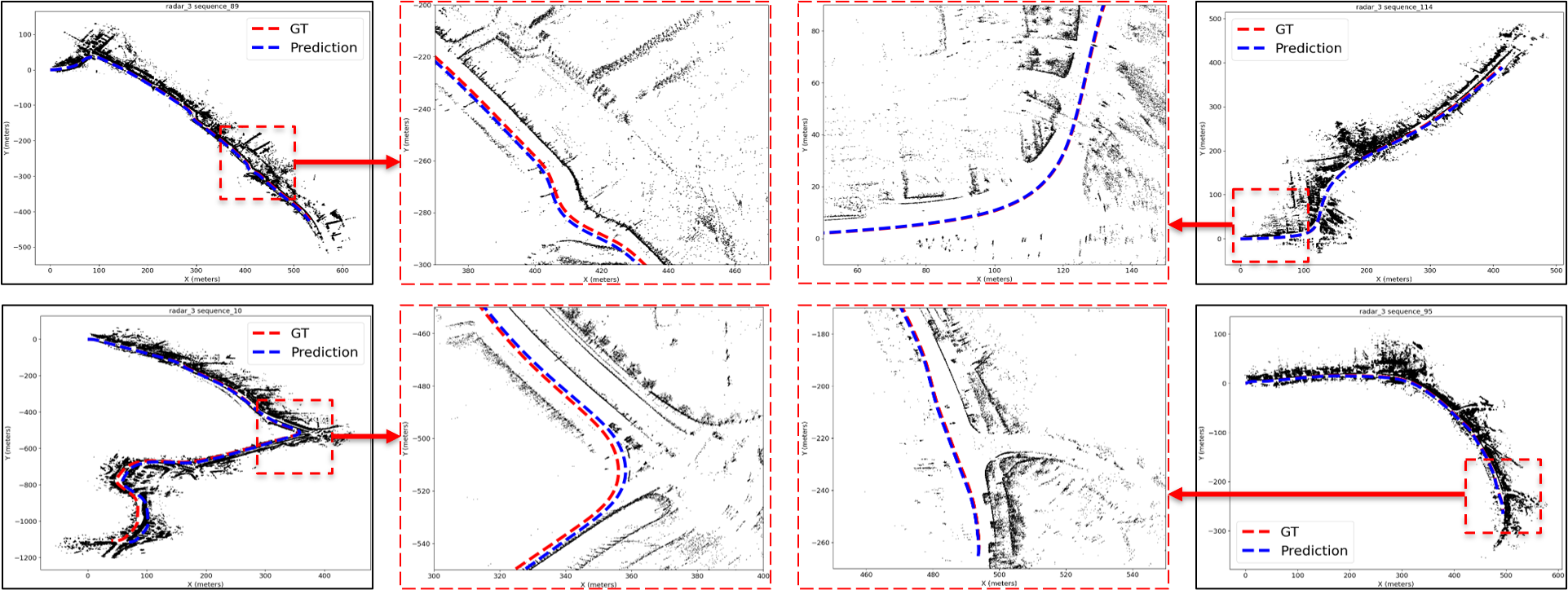}
\caption{Qualitative results of the proposed method for vehicle ego-motion estimation. In this experiment, the proposed method is tested using four sequences from Radar 3, and the estimated ego-motion is converted into vehicle trajectories and displayed on a 2D plane. The red dashed line represents the ground truth trajectory calculated based on the vehicle’s true motion state, and the blue dashed line represents the vehicle's trajectory calculated based on the estimated motion state. The black dots are predicted static objects, accumulated over all radar frames of the tested sequence.}
\label{fig:ego_motion_est}
\end{figure*}

\begin{figure}[!t]
\centering
\includegraphics[width=0.48\textwidth]{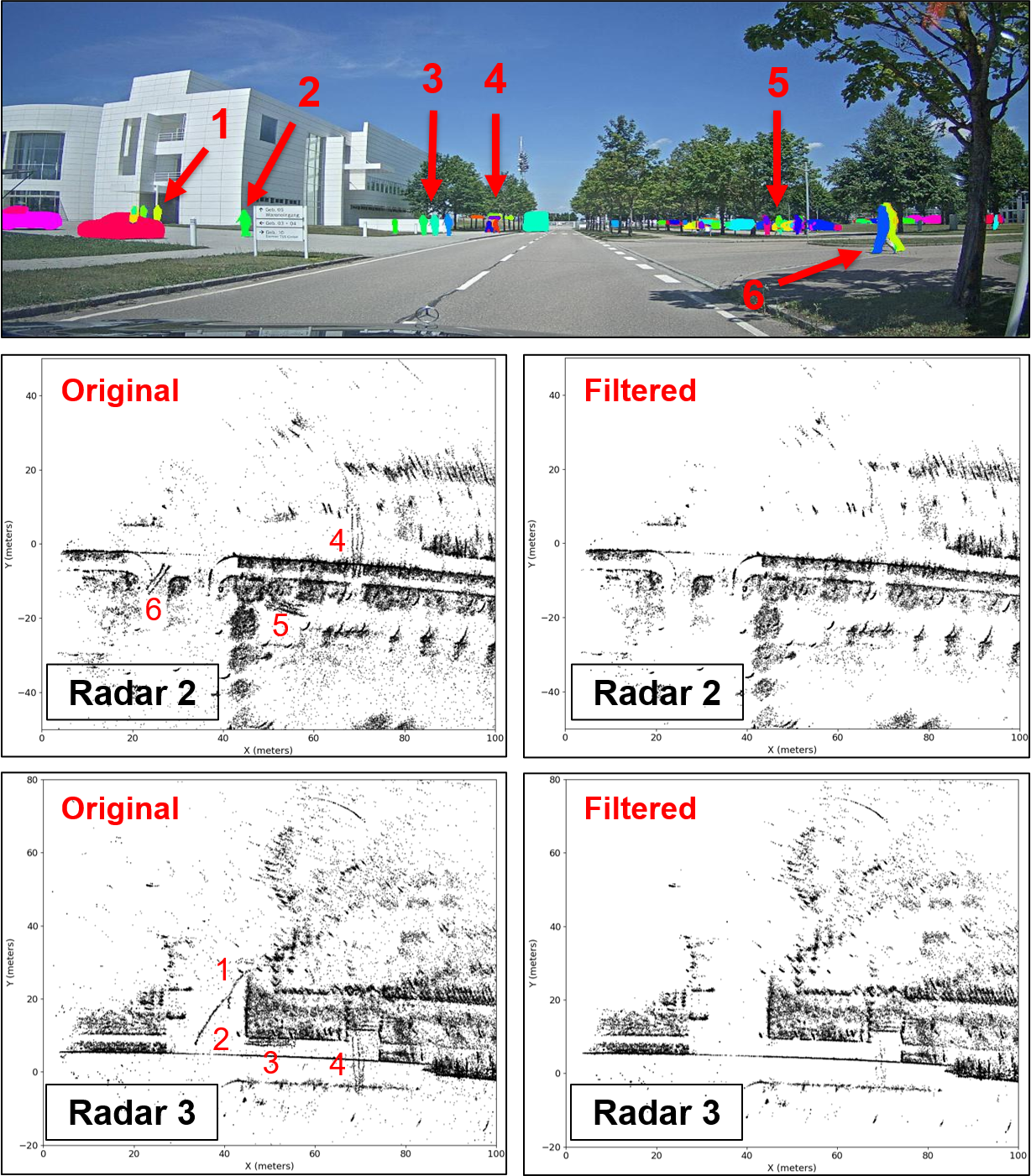}
\caption{Point cloud map constructed using the output of the proposed method. The first row shows the image from the reference camera, in which there are six (groups of) moving pedestrians. The second row shows the point cloud images generated using Radar 2, and the last row is generated using Radar 3. The original map is generated by fusing multiple radar point clouds so that the trajectories of moving objects can be seen. The filtered map is generated using the model's predictions, thus only showing the static environment.}
\label{fig:mapping}
\end{figure}

In addition to segmentation, this method can also simultaneously estimate the 2D motion of the moving ego-vehicle, including forward velocity and rotation rate. Furthermore, by incorporating temporal information, this study can also calculate the vehicle's 2D trajectory, thereby constructing a point cloud map. Figure \ref{fig:ego_motion_est} shows vehicle trajectories calculated from the model’s output on four test sequences from Radar 3. Although in each scene the ego vehicle travels more than 500 meters and the trajectory accumulates errors in the ego-motion estimation, the estimated trajectory still closely follows the GT vehicle trajectory, demonstrating the reliable performance of the proposed method for vehicle localization. Furthermore, thanks to the explicit separation of static objects, the outlines of streets, road edges, and surrounding infrastructure can be clearly seen in the zoomed-in figure, which is of great value for applications such as mapping, drivable road space detection, and semantic segmentation. A more vivid example of using predicted labels for environment mapping is shown in Figure \ref{fig:mapping}, which shows a dynamic environment with six (groups of) walking pedestrians captured by Radar 2 and Radar 3. The trajectories of these moving instances can be clearly observed in the original accumulated radar point cloud. In contrast, filtering the radar point cloud based on the model's prediction removes these trajectories and false positives, leaving behind a distinct outline of the environment. \par

\section{Conclusions}\label{conclusions}
Knowledge of the ego-vehicle velocity and the positions of moving and static objects is sufficient for many radar perception tasks and ensures driving safety, especially in harsh environmental conditions where optical sensors cannot operate. Therefore, unlike traditional radar segmentation research, which requires significant effort to overcome the fundamental limitations of existing radars with limited success, this research reframes the radar segmentation objective as a dual task, which is simpler but more meaningful and reliable for radar data. Specifically, the outcome of this research is a neural network-based solution that can work independently, perform automatic feature extraction, separate moving and static objects, and provide vehicle motion status, all at the same time. According to the literature review, this approach could have a significant impact on radar signal processing, as the authors found that understanding vehicle motion and locating static and moving objects are crucial initial steps in many radar perception tasks. The method has been thoroughly evaluated on the RadarScenes dataset using challenging scenes, novel evaluation metrics, and refined object labels. Results confirm both the feasibility of the dual task using unprocessed radar point clouds and the superior performance of the proposed approach. The network is extremely lightweight (0.15 M parameters) yet achieves high scores in moving object segmentation (IoU = 0.86, F1 = 0.92) and accurate ego-vehicle motion estimation and localization (RTE\_50 = 1.8 m). For future work, extending the approach to estimate and track the velocities of other moving objects beyond the ego-vehicle would be a promising direction. \par

\bibliographystyle{IEEEtran}
\bibliography{reference} 
\vspace{-3 mm}
\section*{Biography Section}
\vspace{-13 mm}
\begin{IEEEbiography}[{\includegraphics[width=1in,height=1.25in,clip,keepaspectratio]{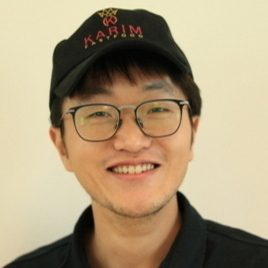}}]{Simin Zhu}
received his BSc degree in Electrical Engineering and Automation from the Central South University in 2016. Afterward, he worked for 1.5 years as a hardware engineer at Huawei Technology Co. Ltd. In 2019, Simin started his master's study at Delft University of Technology (TU Delft). During his master's program, he specialized in radar signal processing and machine learning. In November of 2021, he completed his master's thesis and graduated from the Microwave Sensing, Signals and Systems (MS3) group at TU Delft. In December 2021, he continued his research in the MS3 group as a Ph.D. candidate.
\end{IEEEbiography}\vspace{-11 mm}
\begin{IEEEbiography}[{\includegraphics[width=1in,height=1.25in,clip,keepaspectratio]{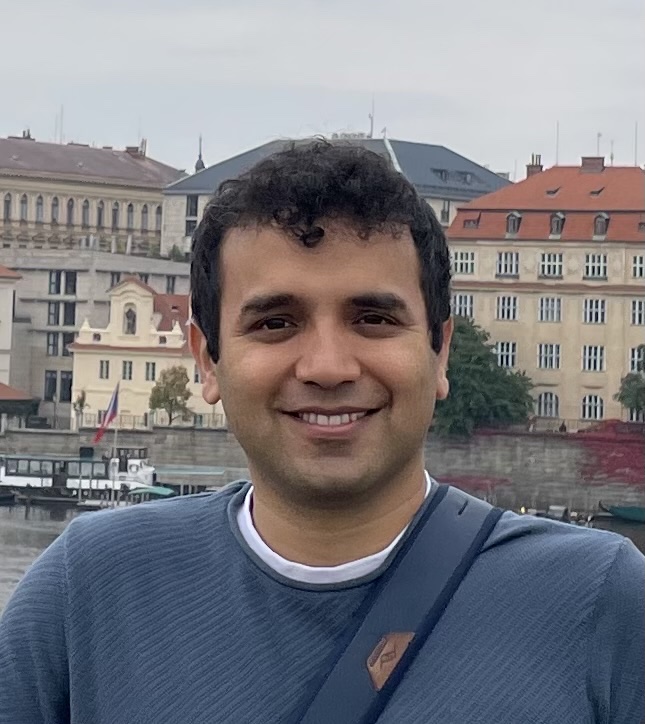}}]{Satish Ravindran}
has around 12 years of experience developing AI solutions for different industries such as Autonomous Driving, Intelligent Traffic Sensing (ITS) and IoT. He has worked on a wide spectrum of applications in AI including NLP, Computer Vision and Radar Processing. He joined NXP in 2018 and is currently the AI Technical Lead for Radar Innovations working in the NXP R\&D division. He has led the development of a comprehensive portfolio of AI applications at all stages of the radar processing chain, from signal processing to perception. He is also helping in the definition of the next generation of NXP SoCs and has multiple patents and papers published in radar signal processing and AI solutions.
\end{IEEEbiography}\vspace{-11 mm}
\begin{IEEEbiography}[{\includegraphics[width=1in,height=1.25in,clip,keepaspectratio]{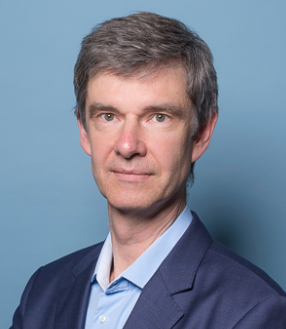}}]{Alexander G. Yarovoy}
(FIEEE’ 2015) graduated from the Kharkov State University, Ukraine, in 1984 with the Diploma with honor in radiophysics and electronics. He received the Candidate Phys. \& Math. Sci. and Doctor Phys. \& Math. Sci. degrees in radiophysics from the same university in 1987 and 1994, respectively.
In 1987 he joined the Department of Radiophysics at the Kharkov State University as a Researcher and became a Full Professor there in 1997. From September 1994 through 1996 he was with Technical University of Ilmenau, Germany as a Visiting Researcher. Since 1999 he is with the Delft University of Technology, the Netherlands. Since 2009 he leads there a chair of Microwave Sensing, Systems and Signals.
His main research interests are in high-resolution radar, microwave imaging and applied electromagnetics (in particular, UWB antennas). He has authored and co-authored more than 600 scientific or technical papers, eleven patents and fourteen book chapters. He is the recipient of the European Microwave Week Radar Award for the paper that best advances the state-of-the-art in radar technology in 2001 (together with L.P. Ligthart and P. van Genderen) and in 2012 (together with T. Savelyev). In 2023 together with Dr. I.Ullmann, N. Kruse, R. Gündel and Dr. F. Fioranelli he got the best paper award at IEEE Sensor Conference. In 2010 together with D. Caratelli Prof. Yarovoy got the best paper award of the Applied Computational Electromagnetic Society (ACES).
In the period 2008-2017 Prof. Yarovoy served as Director of the European Microwave Association (EuMA). He is and has been serving on various editorial boards such as that of the IEEE Transaction on Radar Systems. From 2011 till 2018 he served as an Associated Editor of the International Journal of Microwave and Wireless Technologies. He has been member of numerous conference steering and technical program committees. He served as the General TPC chair of the 2020 European Microwave Week (EuMW’20), as the Chair and TPC chair of the 5th European Radar Conference (EuRAD’08), as well as the Secretary of the 1st European Radar Conference (EuRAD’04). He served also as the co-chair and TPC chair of the Xth International Conference on GPR (GPR2004).
\end{IEEEbiography}\vspace{-11 mm}
\begin{IEEEbiography}[{\includegraphics[width=1in,height=1.25in,clip,keepaspectratio]{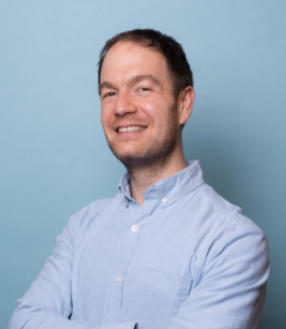}}]{Francesco Fioranelli}
(M'15–SM'19) received the Ph.D. degree with Durham University, Durham, UK, in 2014. He is currently an Associate Professor at TU Delft, The Netherlands, and was an Assistant Professor with the University of Glasgow (2016–2019), and a Research Associate at University College London (2014–2016). 

His research interests include the development of radar systems and automatic classification for human signatures analysis in healthcare and security, drones and UAVs detection and classification, and automotive radar. He has authored over 190 peer-reviewed publications, edited the books on “Micro-Doppler Radar and Its Applications” and "Radar Countermeasures for Unmanned Aerial Vehicles" published by IET-Scitech in 2020, received four best paper awards and the IEEE AESS Fred Nathanson Memorial Radar Award 2024.
\end{IEEEbiography}
\vfill
\end{document}